\documentclass{ws-ijmpa}
\usepackage[super,compress]{cite}
\usepackage{graphicx}
\begin{document}
\markboth{Milan Milo\v sevi\' c, Miika A. Pursiainen, Predrag Jovanovi\'c and Luka \v C. Popovi\'c}{The shape of Fe K$\alpha$ line emitted from\\relativistic accretion disc around AGN black holes}

%%%%%%%%%%%%%%%%%%%%% Publisher's Area please ignore %%%%%%%%%%%%%%%
%
\catchline{}{}{}{}{}
%
%%%%%%%%%%%%%%%%%%%%%%%%%%%%%%%%%%%%%%%%%%%%%%%%%%%%%%%%%%%%%%%%%%%%

\title{The shape of Fe K$\alpha$ line emitted from\\relativistic accretion disc around AGN black holes}

\author{Milan Milo\v sevi\' c}

\address{Faculty of Sciences and Mathematics, University of Ni\v s,\\Vi\v segradska 33, 18000 Ni\v s, Serbia\\
mmilan@seenet-mtp.info}

\author{Miika A. Pursiainen}

\address{Physics and Astronomy, University of Southampton,\\Southampton, SO17 1BJ, UK\\
M.A.Pursiainen@soton.ac.uk}

\author{Predrag Jovanovi\' c$^*$ and Luka \v C. Popovi\'c$^\dagger$}

\address{Astronomical Observatory,\\Volgina 7, 11060 Belgrade, Serbia\\
$^*$pjovanovic@aob.rs\\
$^\dagger$lpopovic@aob.rs}

\maketitle

\begin{history}
\received{Day Month Year}
\revised{Day Month Year}
\end{history}

\begin{abstract}
The relativistically broadened Fe K$\alpha$ line, originating from the 
accretion disc in a vicinity of a super massive black hole, is observed in 
only less than 50\% of type 1 Active Galactic Nuclei (AGN). In this study we 
investigate could this lack of detections be explained by the effects of certain 
parameters of the accretion disc and black hole, such as the inclination, 
the inner and outer radius of disc and emissivity index.  In order to 
determine how these parameters affect the Fe K $\alpha$ line shape, we 
simulated about 60,000 Fe K $\alpha$ line profiles emitted from 
the relativistic disc. 

Based on simulated line profiles, we conclude that the lack of the Fe K$\alpha$ line detection
in type 1 AGN could, be caused by the specific emitting disc parameters, but also by the
limits in the spectral resolution and sensitivity of the X-ray detectors. 

\end{abstract}

\ccode{PACS numbers:}

%\tableofcontents

\section{Introduction}

Active galaxies are galaxies that have a small core of emission embedded at the 
center of an otherwise typical galaxy. This core is typically highly variable 
and very bright compared to the rest of the galaxy. Active galaxies most likely 
represent one phase in galaxy evolution. Their cores, Active Galactic 
Nuclei (AGN), are one of the powerful radiation sources in the 
universe. The luminosity of typical AGN is in the range of $10^8 - 10^{14} L_\odot$. The enormous amount of radiation is coming from an accretion disc that surrounding a Supermassive Black Hole (SMBH) that is supposed to be in the centre of an AGN. 

The structure off all AGN seems to be similar: the central SMBH is surrounded by a 
optically thick and geometrically thin accretion disc that emits in a wide wavelength range from the X-ray to the optical spectral band, mostly in the continuum. The X-ray and UV radiation of the disc is ionizing the gas in so called the Broad Line Region (BLR) that emits broad emission lines. BLR is surrounded by a cold gas in the form of a torus, that emits in the infrared spectral band, and can obscure the BLR (and the accretion disc) emission. Therefore, we observe AGN with the broad lines (unobscured by the torus, so called type 1 AGN), and without broad emission lines (obscured AGN, so called type 2 AGN) \cite{Peterson1997}.

As we noted above, the accretion disc emits mostly in the continuum, but the inner part of the accretion disc (beside the X-ray continuum) emits X-ray lines, among them Fe K$\alpha$ spectral line at 6.4 keV. The line usually has an asymmetric shape with narrow bright blue and wide faint 
red peak. Since this line is produced close to the first marginally stable orbit, it 
is an important indicator of accreting flows around SMBH, as well as of the 
spacetime geometry in these regions \cite{Jovanovic2012, Jovanovic2012a}.

The first results from satellite ASCA showed that Fe K$\alpha$ line is very 
common in spectra of the type 1 AGN and statistical evidence of broadening was 
found in $\sim 75\% $ of sample \cite{Fabian1989, Nandra1997}. However, more 
recent studies of the same type of AGN showed that there is relativistic line 
broadening in only $54\pm 10\% $ of the sample, and only around $30\%$ require 
the line to originate from the vicinity of the SMBH \cite{Nandra2007}.

In this paper we study influence of disc outer radius on the shape of 
Fe K$\alpha$ spectral line for different disc parameters. 

The paper is organized as follows. In Sec. \ref{sec:modeling} 
we present method for modeling the Fe K$\alpha$ 
spectral line profile. In the following Sec. \ref{sec:simulation} we present parameters 
we used in out simulations. In Sec. \ref{sec:results} obtained results are shown 
and discussed. Finally in Sec. \ref{sec:conclusions}, we 
summarize our results and give conclusions.

\section{The Fe K$\alpha$ line and SMBHs of AGN}\label{sec:modeling}
%\subsection{The Fe K$\alpha$ spectral line}

\begin{figure}
\centering
  \includegraphics[width=0.5\linewidth]{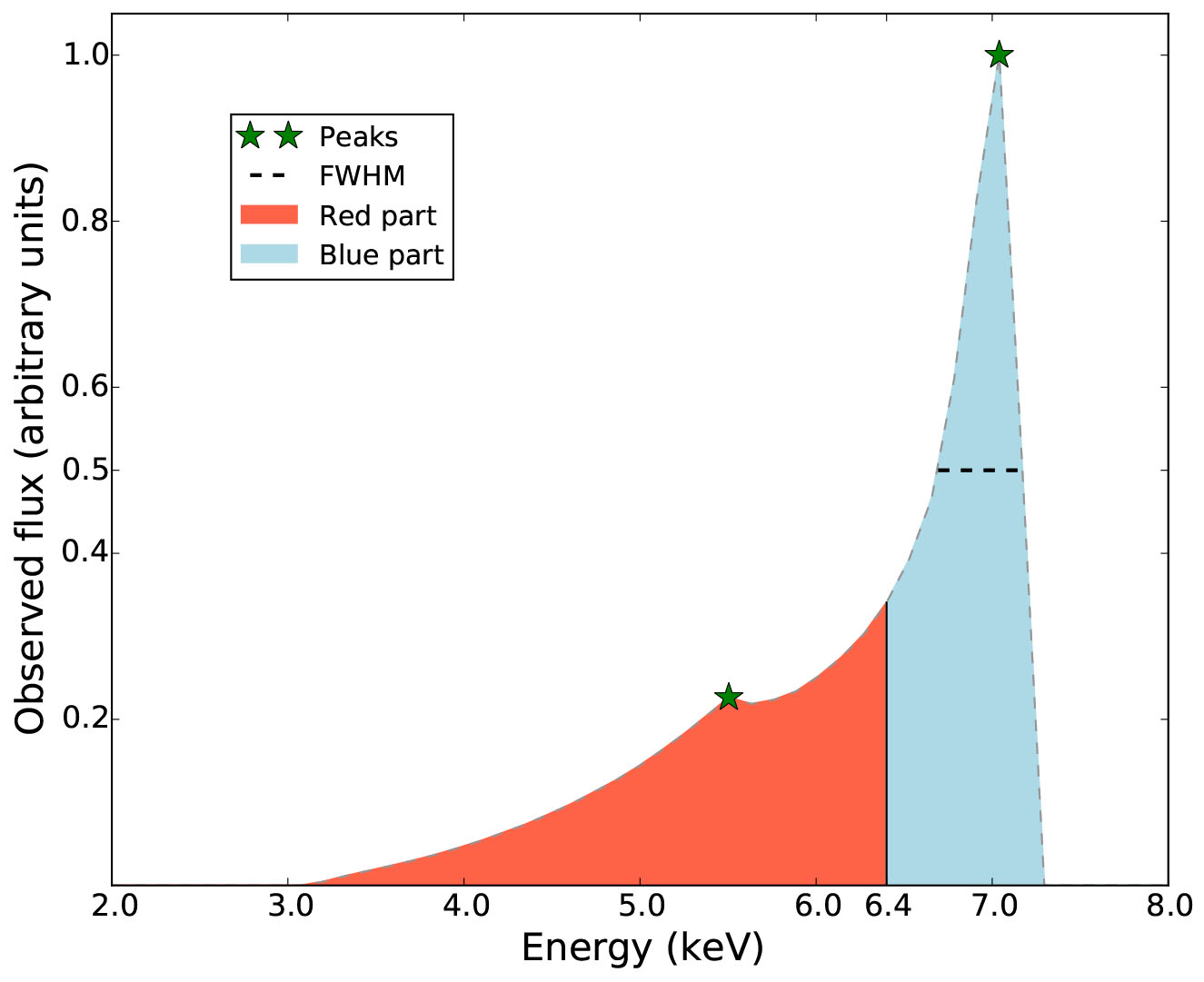}
  \caption{Schematic figure of the calculated parameters of the profile of the spectral line. The asymmetricity ratio was found by dividing the area colored in blue by the area colored red. Dashed black line represents the Full-Widths at Half-Maximum (FWHM).}
  \label{fig:bluered}
\end{figure}

The relativistic component of the Fe K$\alpha$ line was discovered by Tanaka et al. in 1999. They obtained 
the first convincing proof for the existence of the Fe K$\alpha$ line in AGN 
spectra after four-day observations of Seyfert 1 galaxy MCG-6-30-15 
\cite{Tanaka1995}. The Fe K$\alpha$ in this object has a pretty broad profile. If the line 
originated from an arbitrary radius of a nonrelativistic (Keplerian) accretion 
disc it would have a symmetrical profile (due to Doppler effect) with two peaks: 
a ”blue” one which is produced by emitting material from the approaching side of 
the disc in respect to an observer, and a ”red” one which corresponds to 
emitting material from the receding side of the disc (Fig. \ref{fig:bluered}). 
The widest parts of the Fe K$\alpha$ line arise from the innermost regions of 
the disc, where the rotation of emitting material is the fastest. It was found 
that, in case of 14 Seyfert 1 galaxies, Full-Widths at Half-Maximum (FWHM) of 
their Fe K$\alpha$ lines correspond to velocities of $\approx 50,000$ km/s, 
however in some special cases (like Seyfert 1 galaxy MCG-6-30-15) FWHM 
corresponds to the velocity of 30\% of the speed of linght \cite{Nandra2007}. It 
means that in the vicinity of the central black hole, orbital velocities of the 
emitting material are relativistic, causing the enhancement of the Fe K$\alpha$ 
line "blue" peak in regard to its "red" peak.

In the case of the line that originates from a relativistically rotating acration 
disc of an AGN the resulting profile of the Fe K$\alpha$ is a composition of 
three different effects \cite{Jovanovic2012}:
\begin{itemize}
\item Doppler shift due to rotation of emitting material, which is responsible 
for occurrence of two peaks;
\item Special relativistic effect - the relativistic beaming, which is 
responsible for enhancement of the blue peak with respect to the red one;
\item General relativistic effect - the gravitational redshift, which 
is responsible for smearing of the line profile.
\end{itemize}

These characteristics of the observed Fe K$\alpha$ line profiles represent a 
fundamental tool for investigating the plasma conditions and the spacetime 
geometry in the vicinity of the SMBH of AGN.

\begin{figure}
  \includegraphics[width=\linewidth]{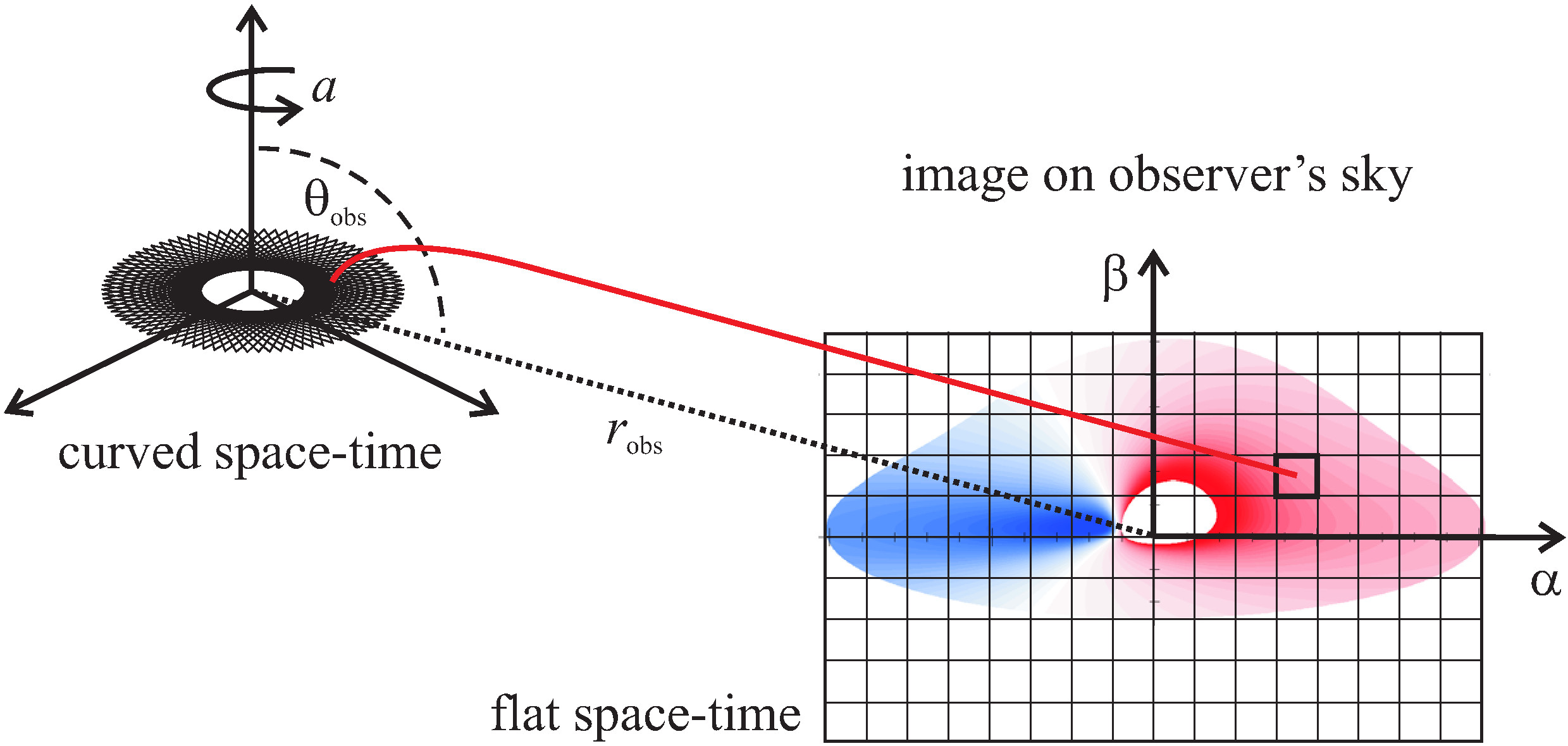}
  \caption{Schematic illustration of the ray-tracing method in the Kerr metric, showing a light ray emitted from some radius of accretion disc around a rotating BH with angular momemntum $a$ and inclination $\theta_{obs}$. The image is visible on observer's sky with coordinates (impact parameters) $\alpha$ and $\beta$. (Figure courtesy: Vesna Borka Jovanovi\' c \cite{Jovanovic2009}) }
  \label{fig:raytrace}
\end{figure}

\section{Numerical simulations}\label{sec:simulation}

The disc emission can be analyzed by numerical simulations taking into account only photon 
trajectories reaching the observer's sky plane. This method is based on so called 
ray-tracing method in Kerr metric \cite{Bao1994,Bromley1997,Fanton1997,Cadez1998}.  The 
image of the disc on the observer's sky is divided into a number of small elements 
(pixels). The color images of the accretion disc which a distant 
observer would see by a high resolution telescope can be obtained in the following way: for each pixel of the image the photon is traced backward from the observer by following the geodesics in a Kerr space-time, until it crosses the plane of the disc. Then, the flux density of the radiation emitted by the disc at that point, as well as the redshift factor of the photon are calculated. The simulated line profiles can be calculated taking into account the intensities and received photon energies of all pixels of the corresponding disc image.

The method used in simulations is based on the pseudo-analytical integration of the geodesic equations which describe the photon trajectories in the general case of a rotating BH having some angular momentum $J$, which gravitational field is therefore described by the Kerr metric  \cite{Cadez1998, Jovanovic2009}:
\begin{equation}
ds^2=-\left(1-{\dfrac{2Mr}{\Sigma}}\right)dt^2
-{\dfrac{4Mar}{\Sigma}}\sin^2{\theta}dt d\phi
+{\dfrac{A}{\Sigma}}\sin^2{\theta}d\phi^2
+{\dfrac{\Sigma}{\Delta}}dr^2+\Sigma d\theta^2,
\label{eq32_1}
\end{equation}
where $(r,\theta,\phi,t)$ are the usual Boyer-Lindquist coordinates, with $c=G=1$ and 
$\Sigma=r^2+a^2\cos^2{\theta}$, $\Delta=r^2+a^2-2Mr$, and 
$A=(r^2+a^2)^2-a^2\Delta\sin^2{\theta}$.

The Kerr metric depends on the angular momentum normalized to the mass $M$ of 
black hole: $a=J/Mc$, $0\le a \le M$.

A photon trajectory in the Kerr metric can be described by three constants of 
motion (the energy at infinity and two constants related to the angular 
momentum, respectively) which, in natural units $c=G=M=1$, have the following 
forms \cite{Cadez1998, Jovanovic2009}: 
\begin{equation}
E=-p_t,\quad \Lambda=p_\phi,\quad Q=p^2_\theta-a^2 E^2 cos^2\theta+\Lambda^2 cot^2\theta,
\end{equation}
where $p$ is the 4-momentum.

Now, two dimensionless parameters $\lambda=\Lambda/E$ and $q=Q^{1/2}/E$ can be 
introdced to express the trajectory of the photon, because it is independent on 
energy of the photon. Parameters $\lambda$ and $q$ are related to the two impact 
parameters $\alpha$ and $\beta$ which describe the apparent position on the 
observer's celestial sphere: 
\begin{equation}
\alpha = -{\dfrac{{\lambda}} {{\sin \theta _{obs}}} }, \qquad
\beta = \pm \left( {q^{2} + a^{2}\cos ^{2}\theta _{obs} - \lambda ^{2}\cot ^{2}\theta _{obs}} \right)^{{\frac{{1}}{{2}}}},
\end{equation}
where the sign of $\beta$ is determined by $\left( {{\dfrac{{dr}}{{d\theta}} 
}}\right)_{obs}$.

The solution of integral equation \cite{Cadez1998}:
\begin{equation}
\pm \int\limits_{r_{em}} ^{\infty}  {{\dfrac{{dr}}{{\sqrt {R\left({r,\lambda ,q} \right)}}} }}  = \pm \int\limits_{\theta _{em}} ^{\theta_{obs}}  {{\dfrac{{d\theta}} {{\sqrt {\Theta \left( {\theta ,\lambda ,q} \right)}}} }},
\label{eq53_1}
\end{equation}
\begin{equation}
\begin{array}{c}
R\left( {r,\lambda ,q} \right) = \left( {r^{2} + a^{2} - a\lambda} \right)^{2} - \Delta {\left[ {\left( {\lambda - a} \right)^{2} + q^{2}} \right]}, \\
\Theta \left( {\theta ,\lambda ,q} \right) = q^{2} + a^{2}\cos ^{2}\theta -\lambda ^{2}\cot ^{2}\theta .
\end{array}
\label{eq53_2}
\end{equation}
provide the photon trajectories (null geodesics) which originate in the accretion disc at some emission radius $r_{em}$ and reach the observer at 
infinity. The integral Equation (\ref{eq53_1}) can be solved in terms of Jacobian elliptic functions, and therefore it is a pseudo-analytical integration. 

Photons emitted at frequency $\nu_{em}$ will reach infinity at frequency $\nu_{obs}$ because of relativistc effects. Their ratio $g = \dfrac{{\nu _{obs}}}
{{\nu _{em}}}$ determines the shift due to these effects. The total observed flux at the observed energy $E_{obs}$ is given by \cite{Fanton1997}:
\begin{equation}
F_{obs} \left( {E_{obs}}  \right) = {\int\limits_{image} {\varepsilon \left({r} \right)}} g^{4}\delta \left( {E_{obs} - gE_{0}}  \right)d\Xi ,
\label{eq53_3}
\end{equation}
where $\varepsilon \left( {r} \right)$ is the disc emissivity, $d\Xi$ is the 
solid angle subtended by the disc in the observer's sky and $E_{0}$ is the rest 
energy.

Image of a simulated accretion disc is obtained in the following way \cite{Jovanovic2009}
\begin{enumerate}
\item values of the input parameters are specified: inner ($R_{in}$) and outer ($R_{out}$)
radii of the disc, angular momentum $a$ of the central 
BH, disc inclination (observer's viewing angle) $\theta_{obs}$ (also, 
denoted by $i$) and parameters defining the disc emissivity
\item constants of motion $\lambda$ and $q$ are calculated for each pair of impact parameters 
$\alpha$ and $\beta$ (i.e. for each pixel on imaginary observer's photographic plate)
\item geodesic Equation (\ref{eq53_1}) is integrated for each pair of $\lambda$ 
and $q$
\item values of shift due to relativistic effects $g$ and observed flux 
$F_{obs}$ are calculated 
\item pixels on imaginary observer's photographic plate are colored according 
to the value of shift $g$ and a simulated disc image is obtained.
\end{enumerate}

\begin{figure}
  \includegraphics[width=\linewidth]{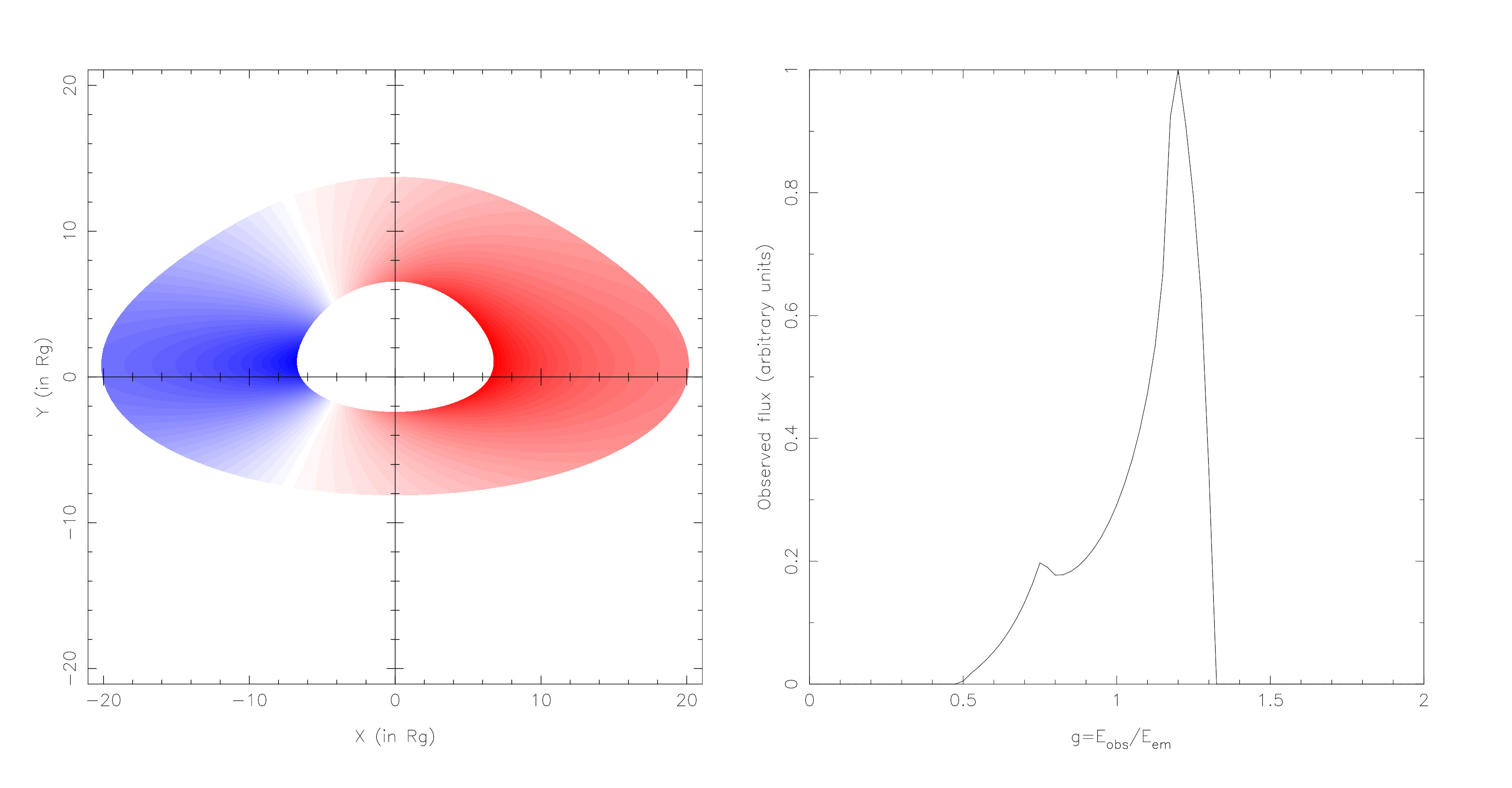}
  \caption{The illustration of simulated an acretion disc (left) and the 
corresponding Fe K$\alpha$ line profile (right). Parameters for simulation are 
$q=2.5$, $i=65$, $R_{in}=r_{ms}$, $R_{out}=20$, $a=0.05$, $nres=5000$ and 
$nbin=80$.}
  \label{fig:lineprofile}
\end{figure}

From the corresponding disc images the simulated line profiles can be calculated by binning the observed flux at all pixels over the bins of shift $g$. In left panels of Fig. \ref{fig:lineprofile} the examples of simulated disc images obtained in such way are presented. The corresponding simulated line profiles are presented in the right panels of the same figure.

\subsection{Disc parameters}

All simulated line profiles are done using the ray-tracing method discribed in 
previous section and proposed by A. Čadež et al. \cite{Cadez1998}. About 60,000 
accretion discs and corresponding Fe K$\alpha$ lines were simulated for various 
set of parameters (Table \ref{tab:params}). We varied values of the emissivity 
index $q$, the inclination $i$, the outer radius $R_{out}$ of the disc and the 
spin $a$ of BH.

The emissivity index $q$ defines the emissivity profile of the 
disc with radius $R$ according to the law $\epsilon(R) \propto R^{-q}$. 
Inclination ranges from $5^{\circ}$ to $80^{\circ}$ and the spin of the BH from 
almost non-rotating ($a=0.05$) up to maximally rotating Kerr BH ($a=0.998$). The 
inner radius $R_{in}$ was determined as the innermost stable orbit around the 
SMBH, also known as \textit{the marginally stable orbit}, $r_\textrm{ms}$. The 
values are $1.24R_\textrm{g}$ for $a=0.998$ and $5.84R_\textrm{g}$ for $a=0.05$.

\begin{table}[ph]
\centering
\tbl{The parameter ranges of the simulated accretion discs.}{
\begin{tabular}{c|c|c}
Parameter								&Values					&Description\\
\hline
$q$ 										&2, 2.5, 3, 4			&Emissivity 
indices\\
$i$ ($^{\circ}$)							&5-80 (5)				
&Inclinations\\
$R_{\textrm{in}}$ ($R_{\textrm{g}}$)		&$r_\textrm{ms}$			&Inner 
disc radii\\
$R_{\textrm{out}}$  ($R_{\textrm{g}}$)   &10, 20, 30, 50, 70, 100	&Outer disc 
radii\\
$a$										&0.05-0.998 (0.1)		&The BH spins 
with step 0.1\\
$nres$									&1000, 3000, 5000		&Number of 
bins\\
$nbin$									&50, 70, 80, 100			&number of 
photons\\
No.										&55296					&The total 
number of simulations 
\end{tabular} \label{tab:params}}
\end{table}

\clearpage

\begin{figure*}[ht!]
\centering
\includegraphics[width=\textwidth]{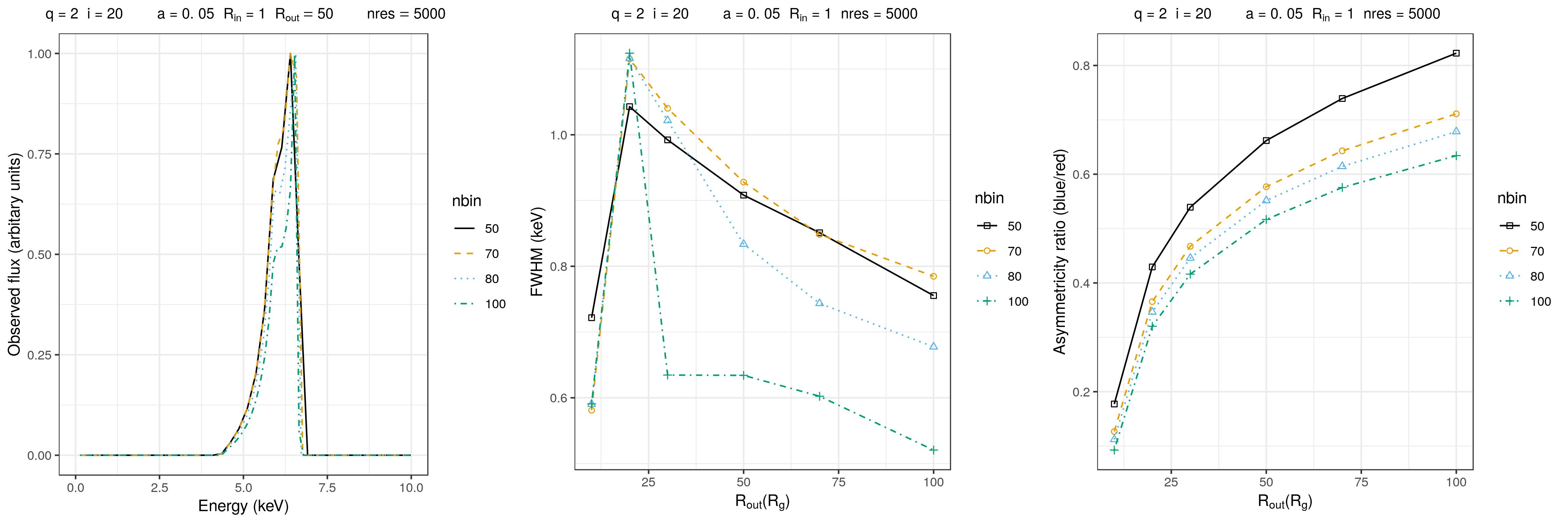} \\
\includegraphics[width=\textwidth]{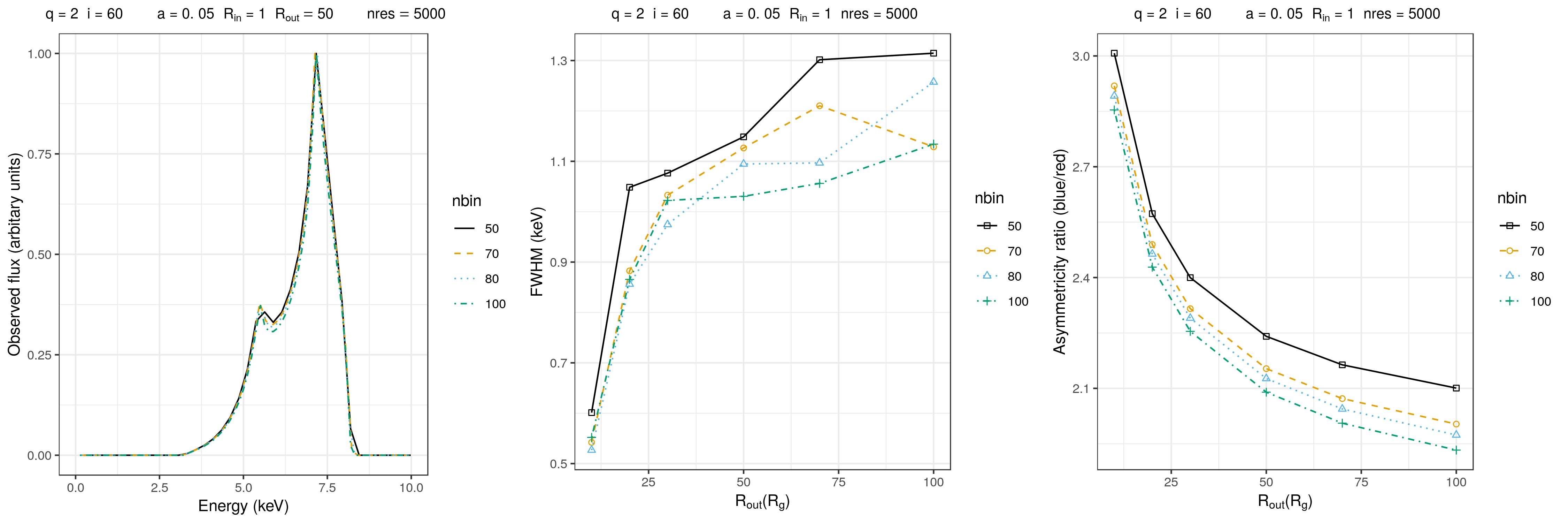} \\
\includegraphics[width=\textwidth]{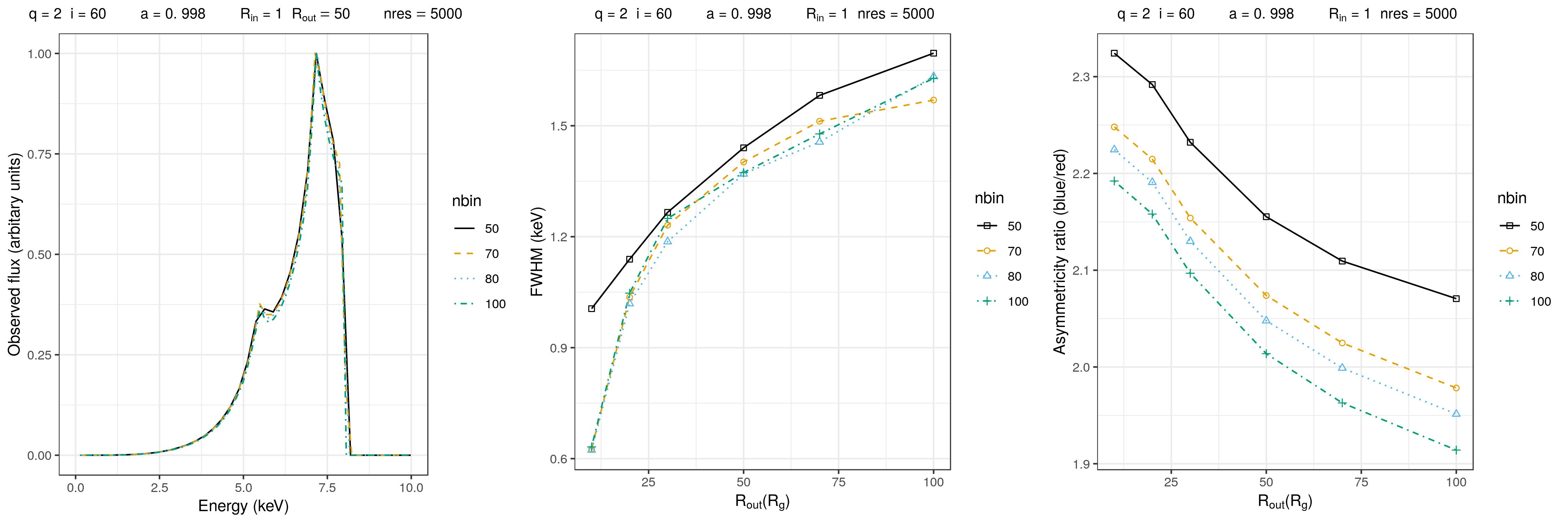} \\
\includegraphics[width=\textwidth]{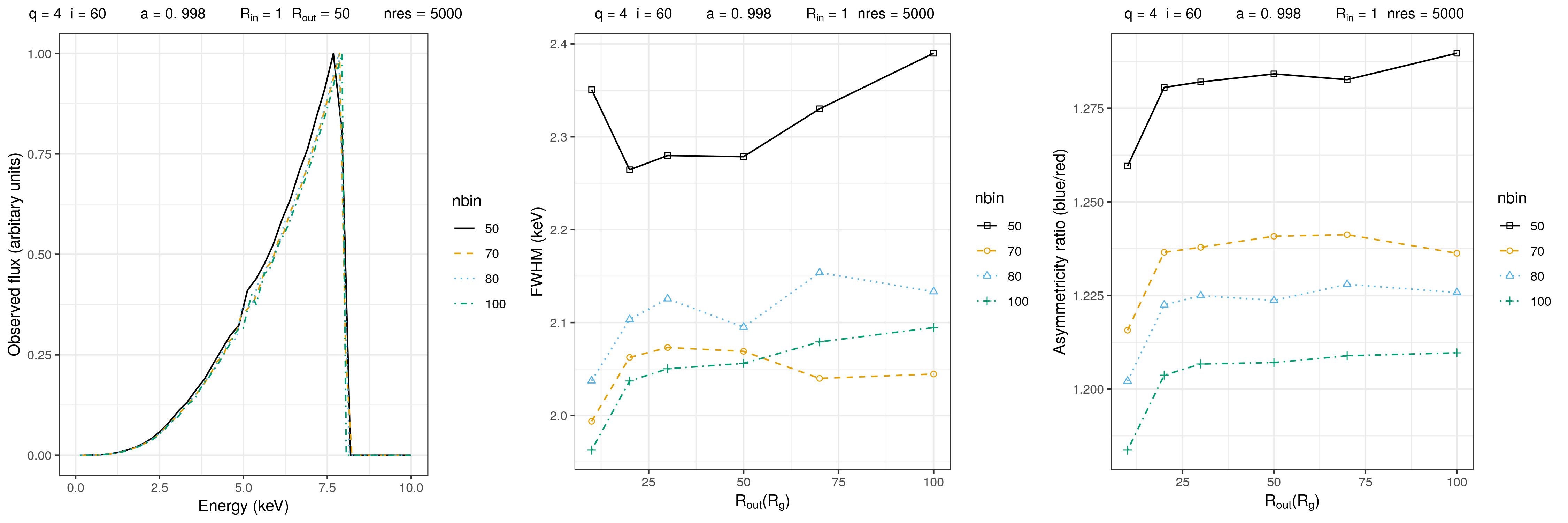} \\
\caption{Influence of number of line bins ($nbin$) on the simulated line profiles (left panels),
its FWHM (middle panels) and asymmetricity ratio (right panels). The presented
results correspond to two different disc inclinations: $i=20^\circ$ and $i=60^\circ$,
and power law emissivity indices: $q=2$ and $q=4$.}
\label{nbin}
\end{figure*}

\clearpage

\begin{figure*}[ht!]
\centering
\includegraphics[width=\textwidth]{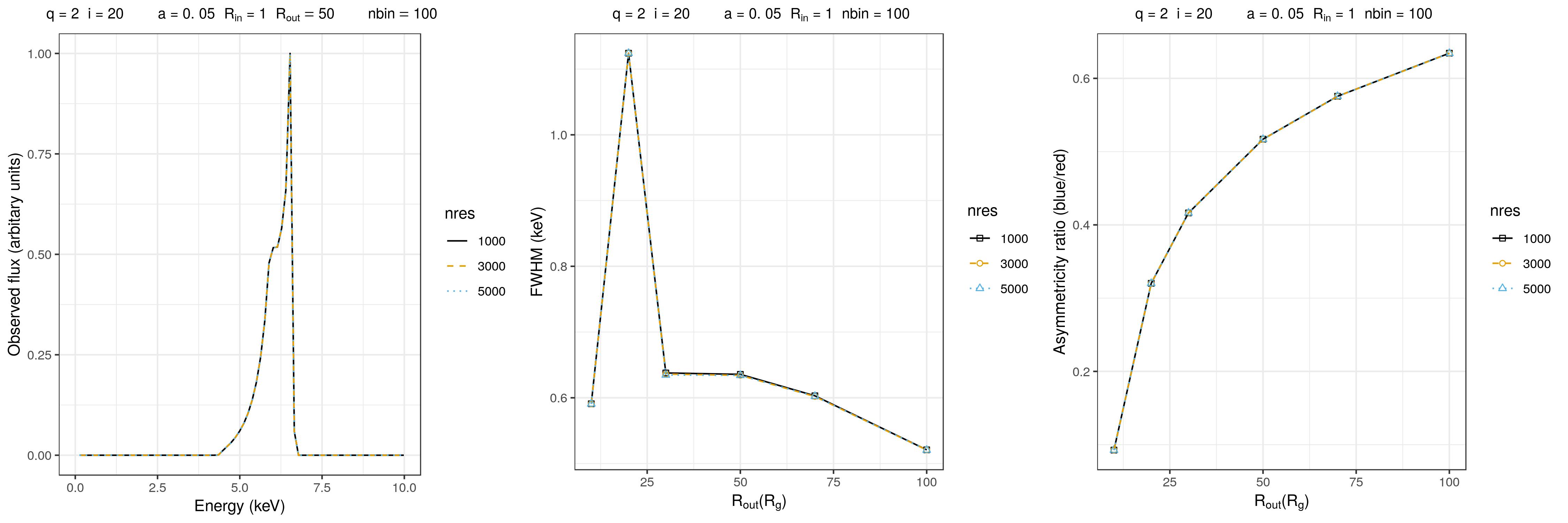} \\
\includegraphics[width=\textwidth]{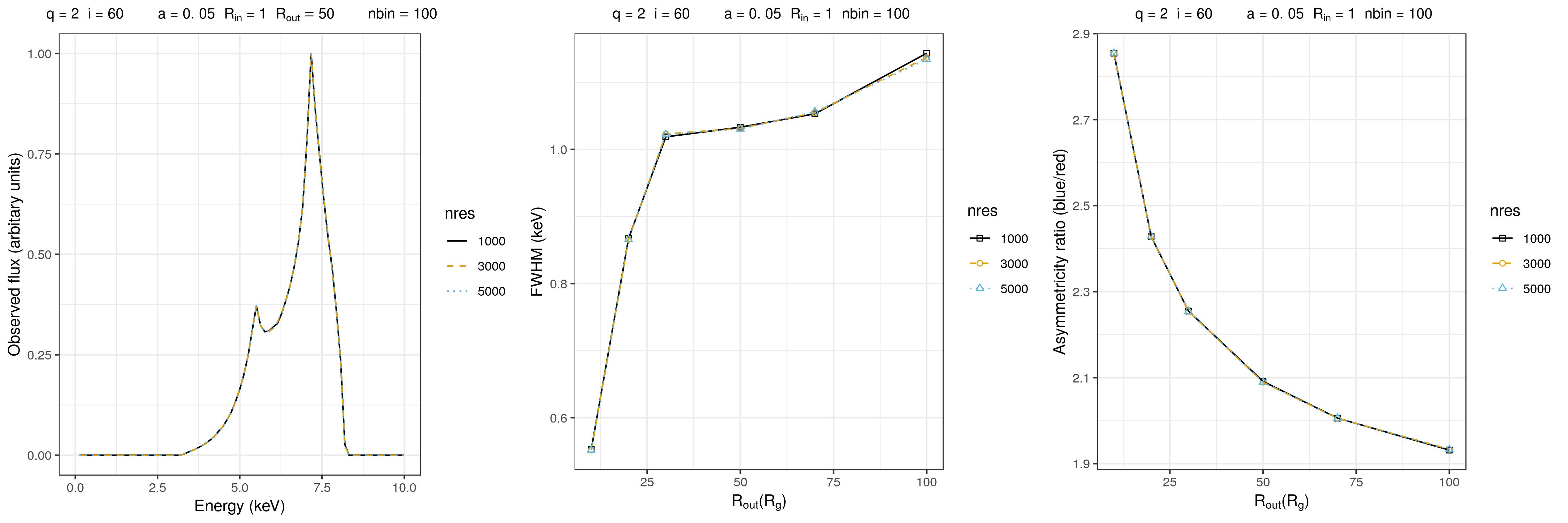} \\
\includegraphics[width=\textwidth]{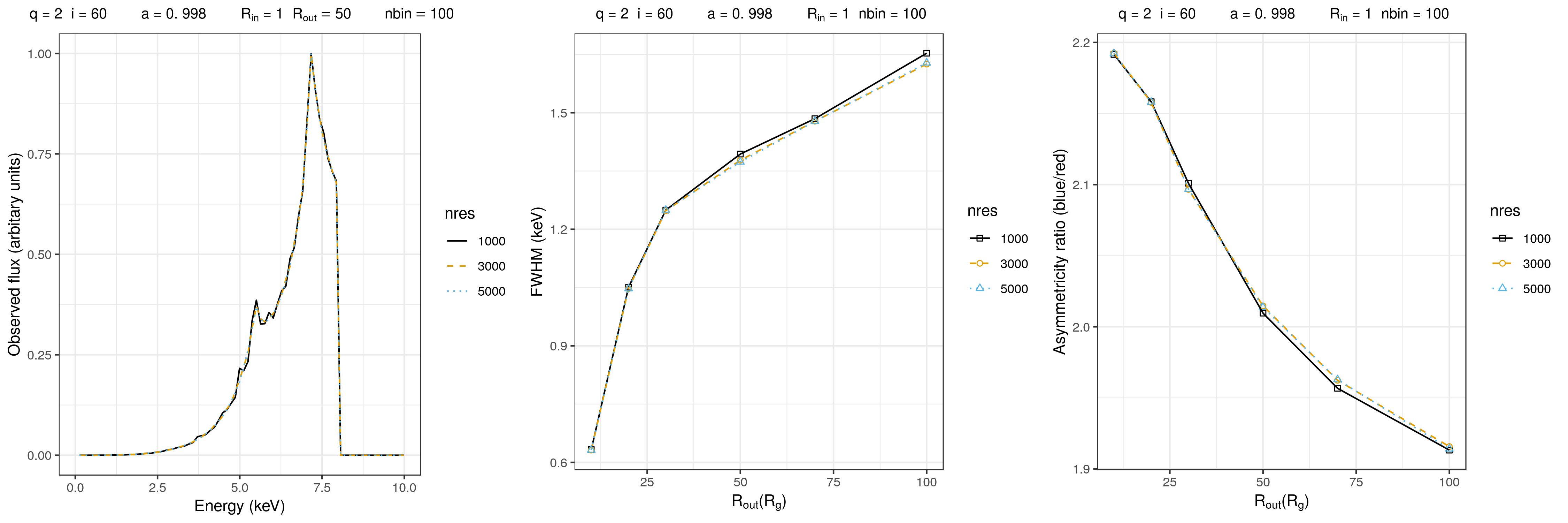} \\
\includegraphics[width=\textwidth]{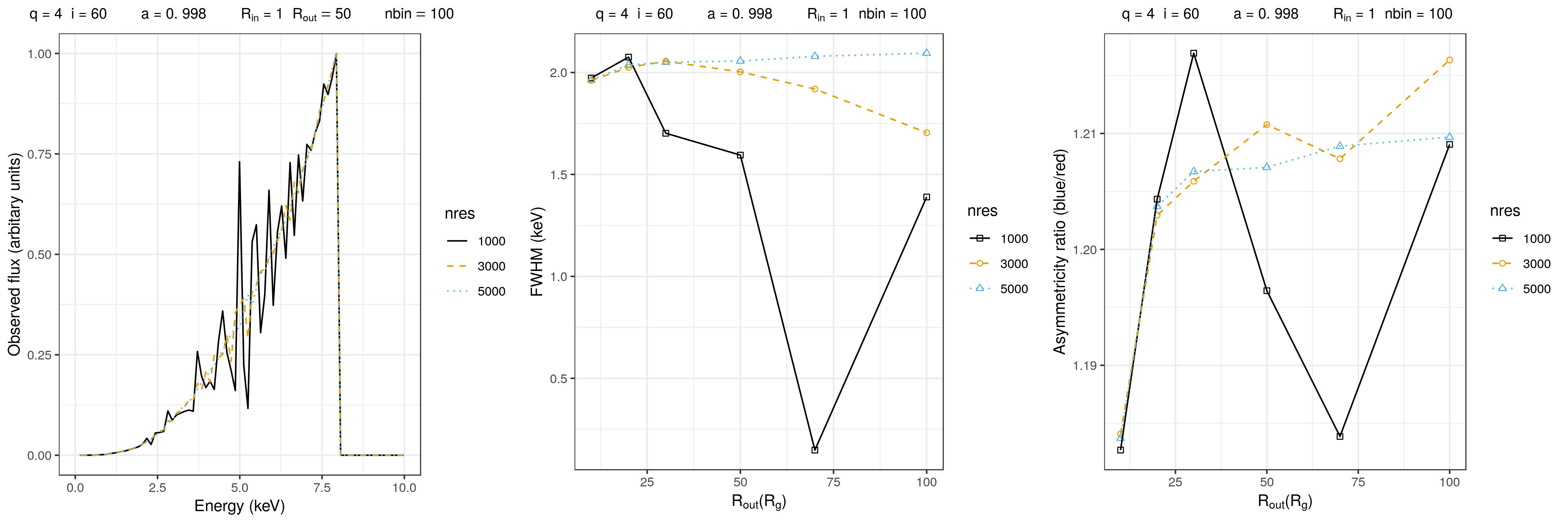} \\
\caption{Influence of number of photons ($nres\times nres$) on the simulated line profiles
(left panels), its FWHM (middle panels) and asymmetricity ratio (right panels).
The presented results correspond to two different disc inclinations: $i=20^\circ$ and
$i=60^\circ$, and power law emissivity indices: $q=2$ and $q=4$.}
\label{nres}
\end{figure*}

\clearpage

\begin{figure*}[ht!]
\centering
\includegraphics[width=\textwidth]{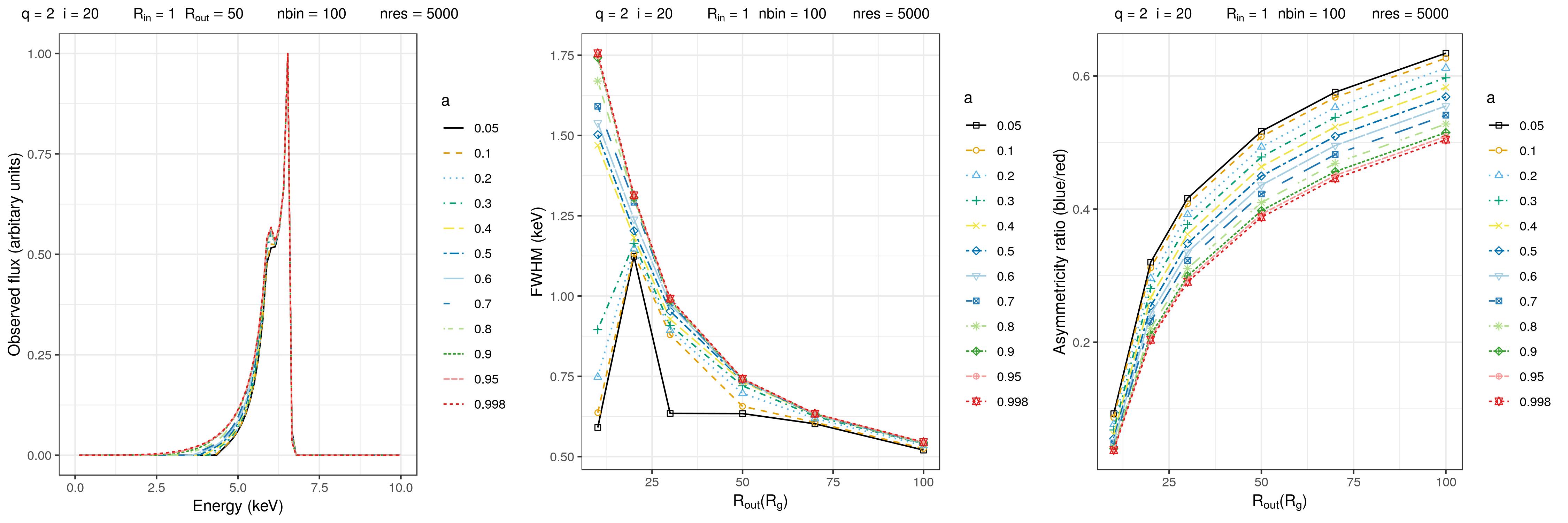} \\
\includegraphics[width=\textwidth]{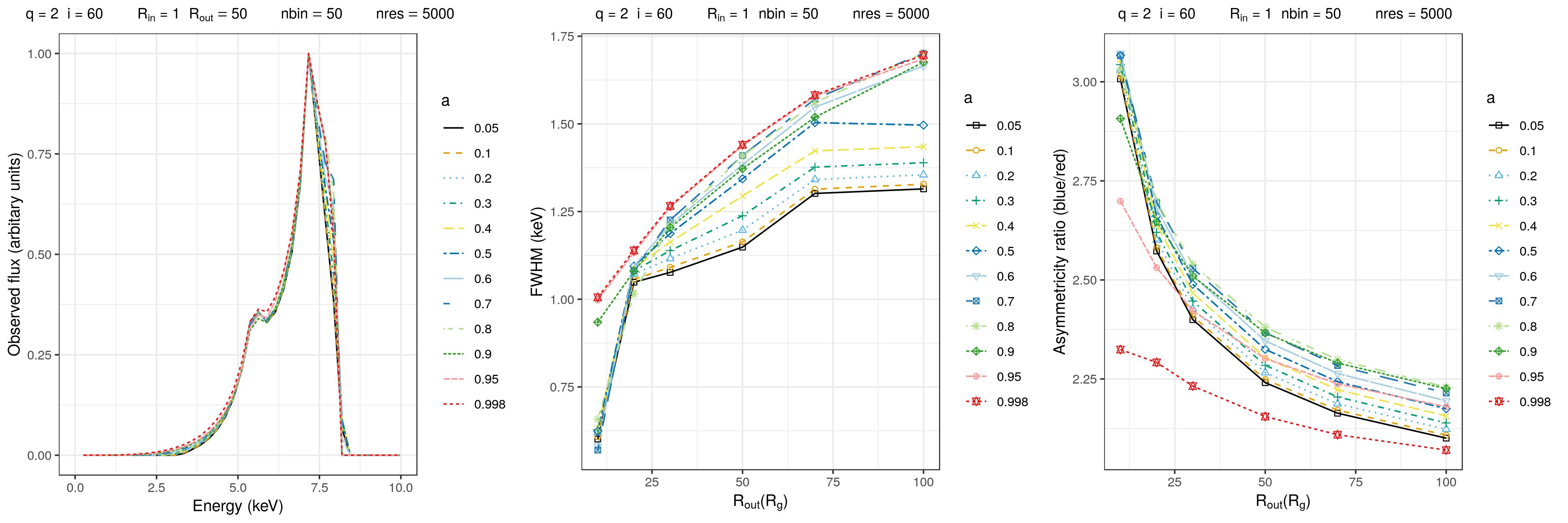} \\
\includegraphics[width=\textwidth]{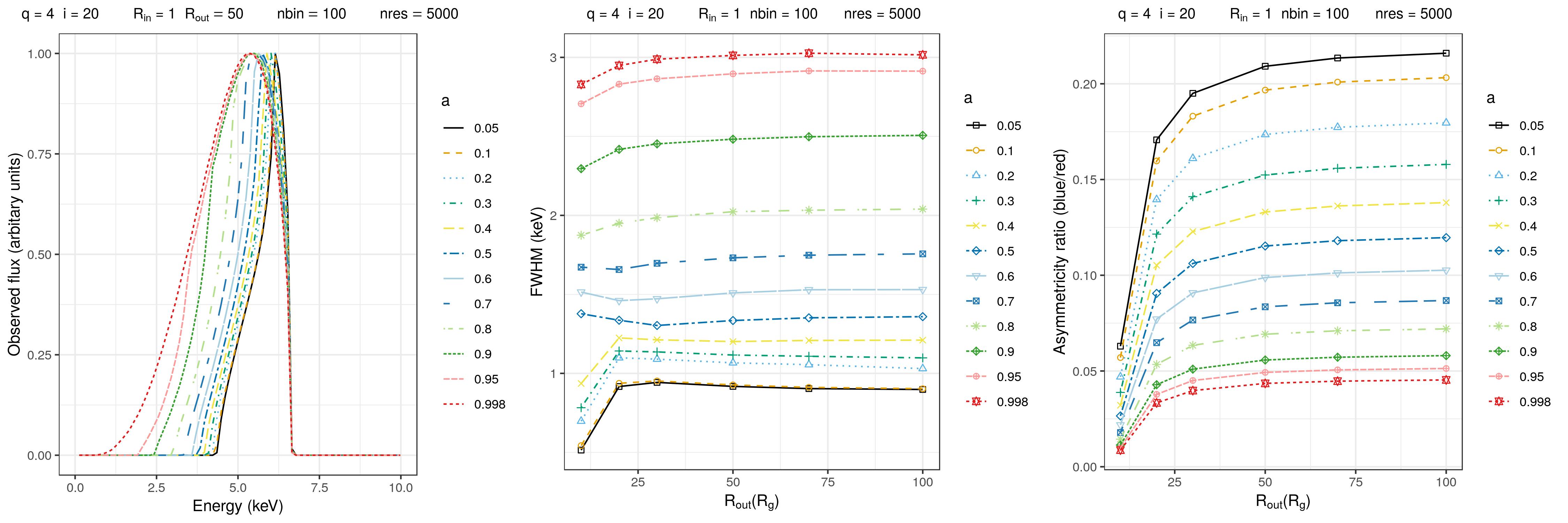} \\
\includegraphics[width=\textwidth]{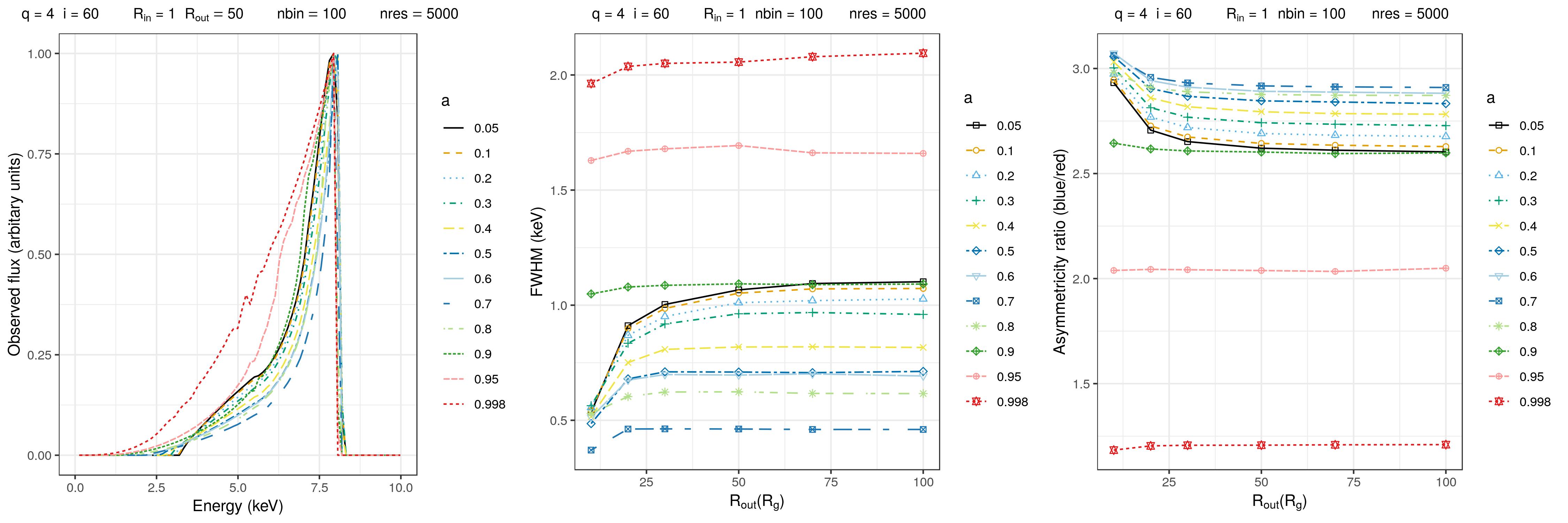} \\
\caption{Influence of SMBH spin $a$ on the simulated line profiles (left panels),
its FWHM (middle panels) and asymmetricity ratio (right panels). The presented
results correspond to two different disc inclinations: $i=20^\circ$ and $i=60^\circ$,
and power law emissivity indices: $q=2$ and $q=4$.}
\label{spin}
\end{figure*}

\clearpage

\begin{figure*}[ht!]
\centering
\includegraphics[width=\textwidth]{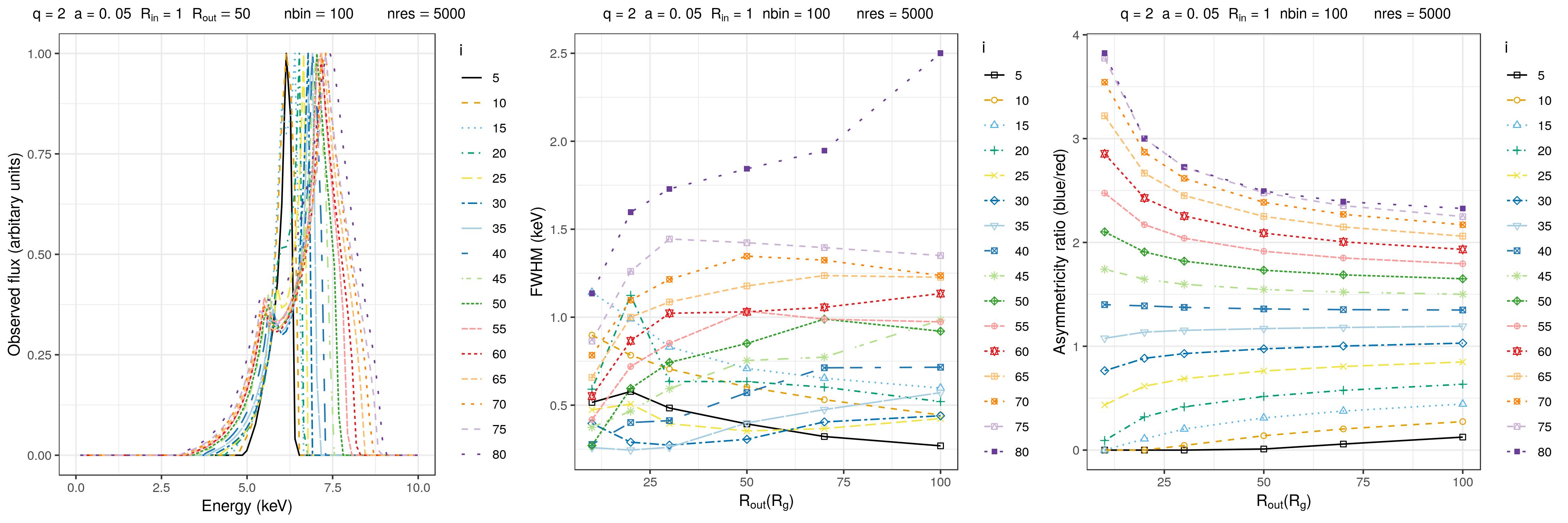} \\
\includegraphics[width=\textwidth]{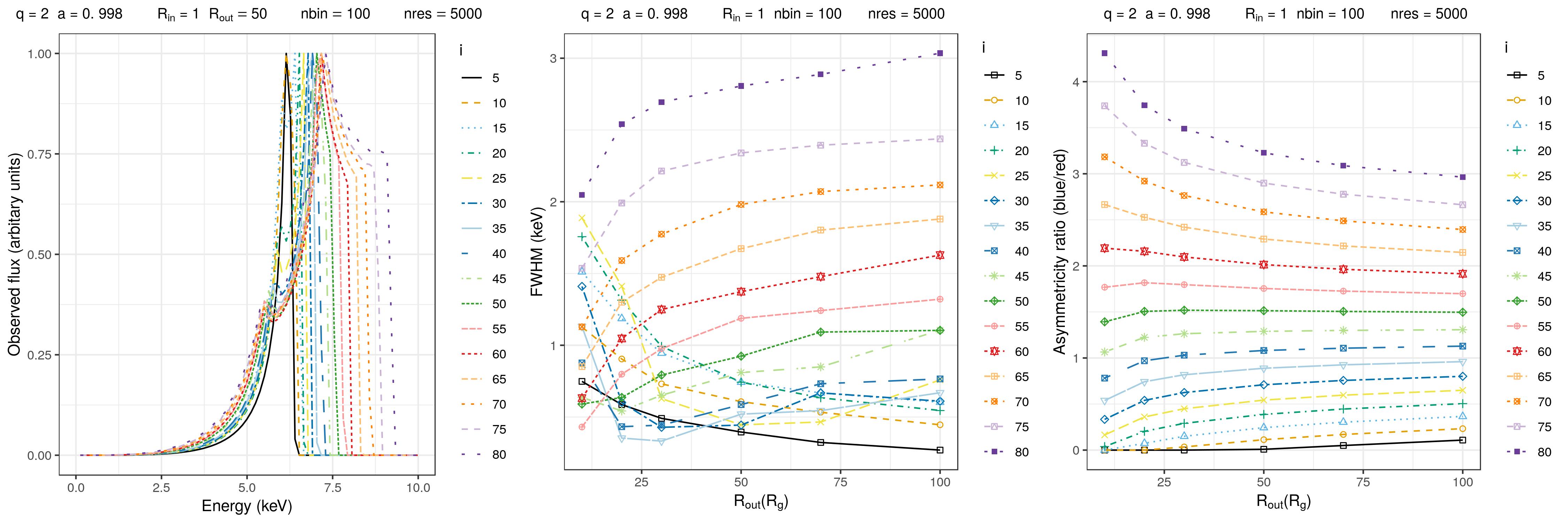} \\
\includegraphics[width=\textwidth]{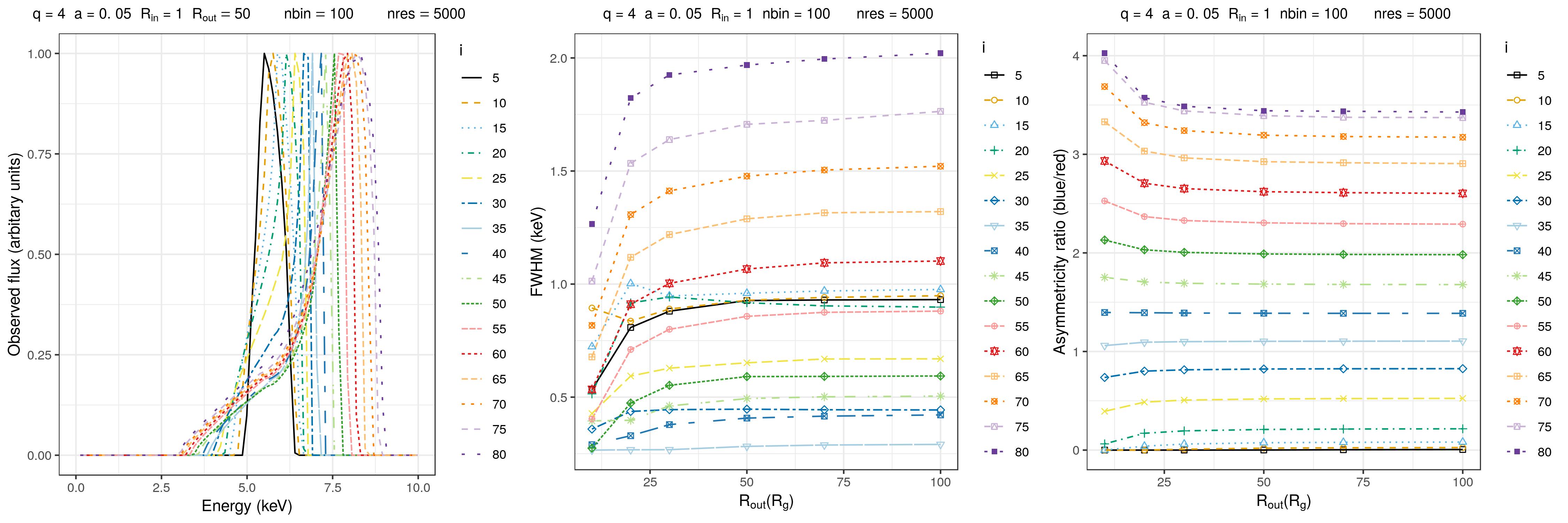} \\
\includegraphics[width=\textwidth]{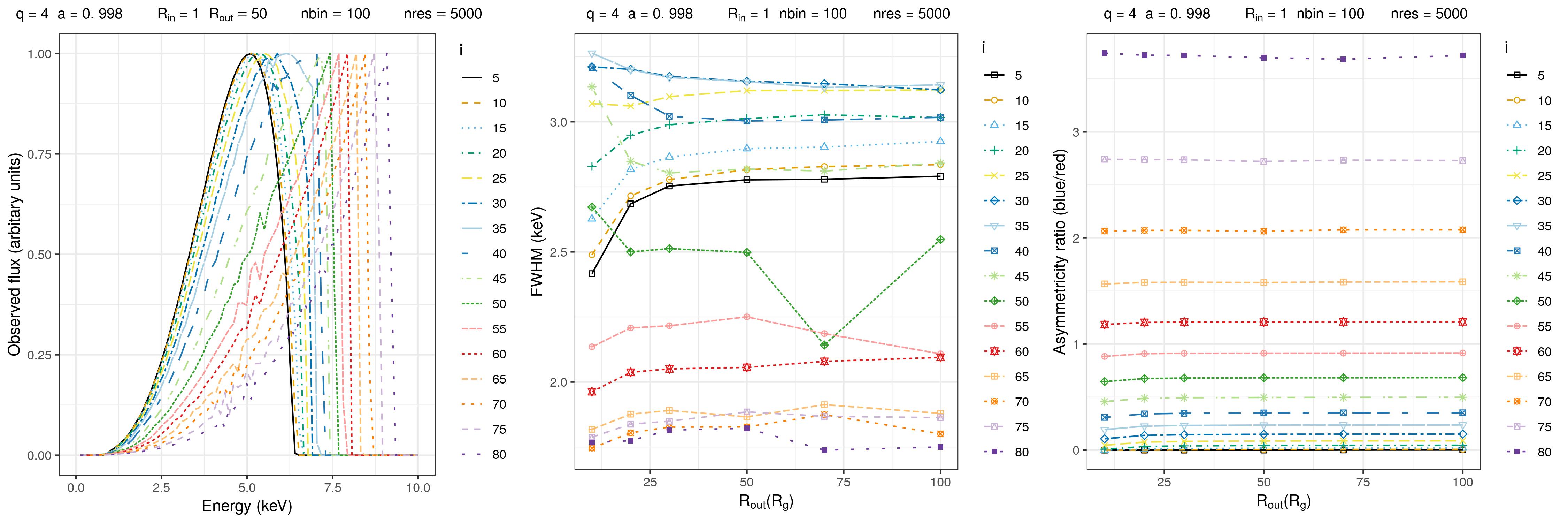} \\
\caption{Influence of disc inclination $i$ (i.e. viewing angle $\theta_{obs}$) on the simulated
line profiles (left panels), its FWHM (middle panels) and asymmetricity ratio (right panels).
The presented results correspond to two different SMBH spins: $a=0.005$ (practically non-rotating
Schwarzschild SMBH) and $a=0.998$ (extremely rotating Kerr SMBH), and power law emissivity indices:
$q=2$ and $q=4$.}
\label{incl}
\end{figure*}

\clearpage

\begin{figure*}[ht!]
\centering
\includegraphics[width=\textwidth]{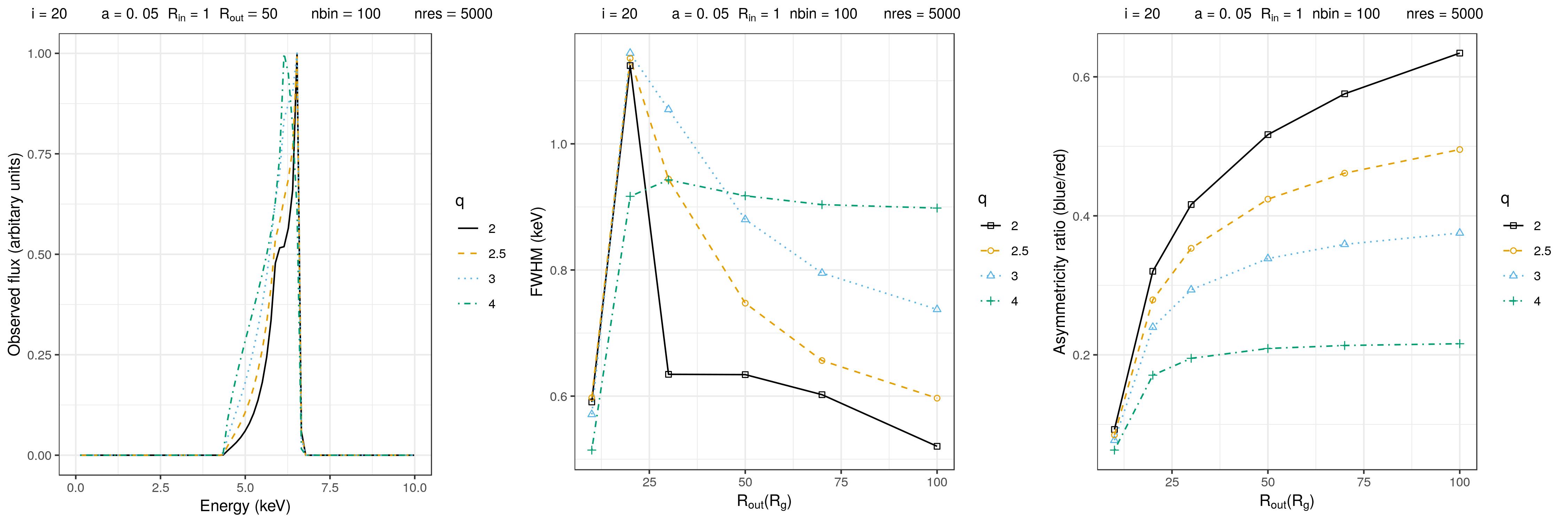} \\
\includegraphics[width=\textwidth]{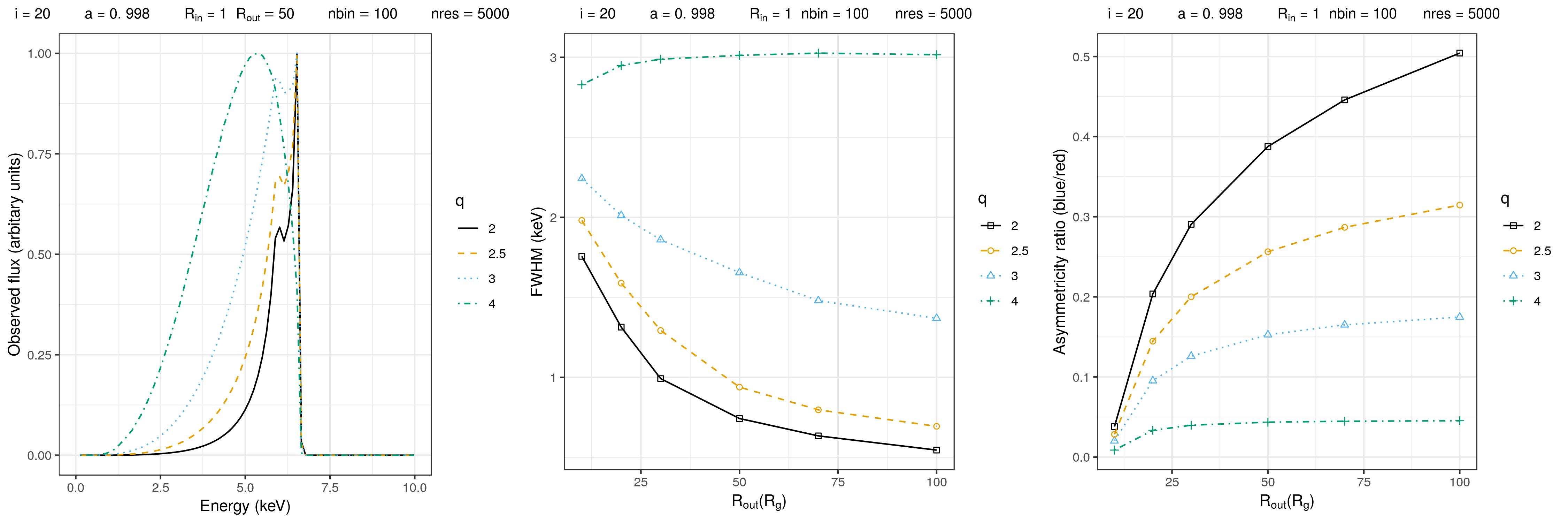} \\
\includegraphics[width=\textwidth]{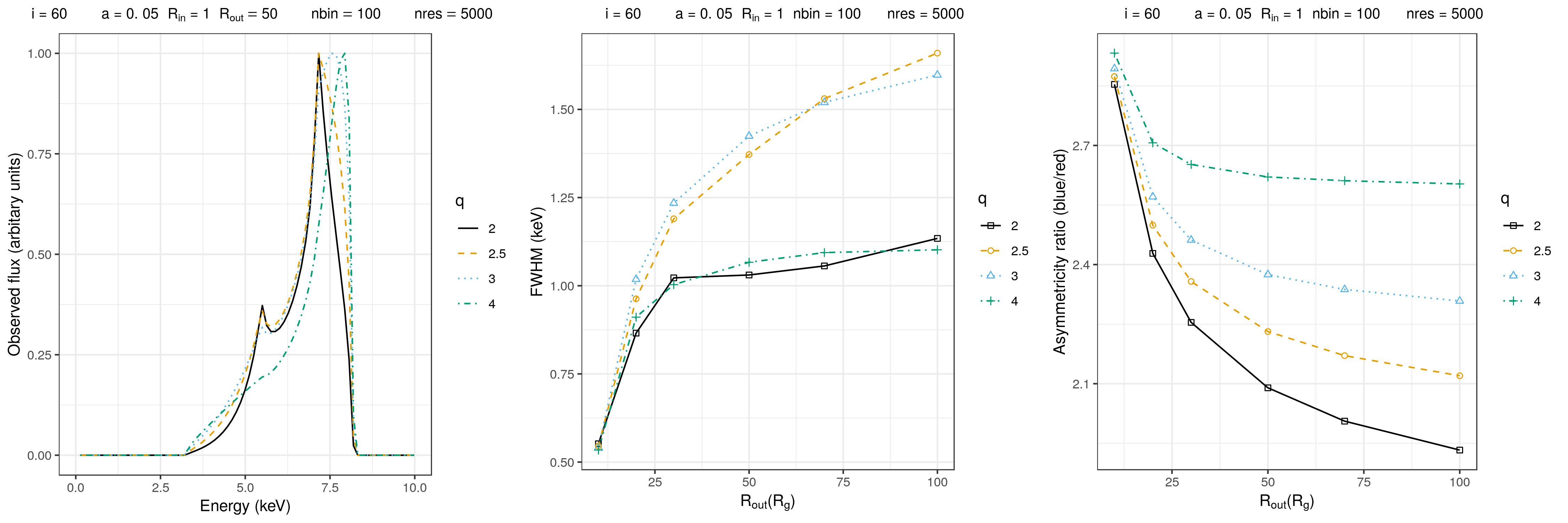} \\
\includegraphics[width=\textwidth]{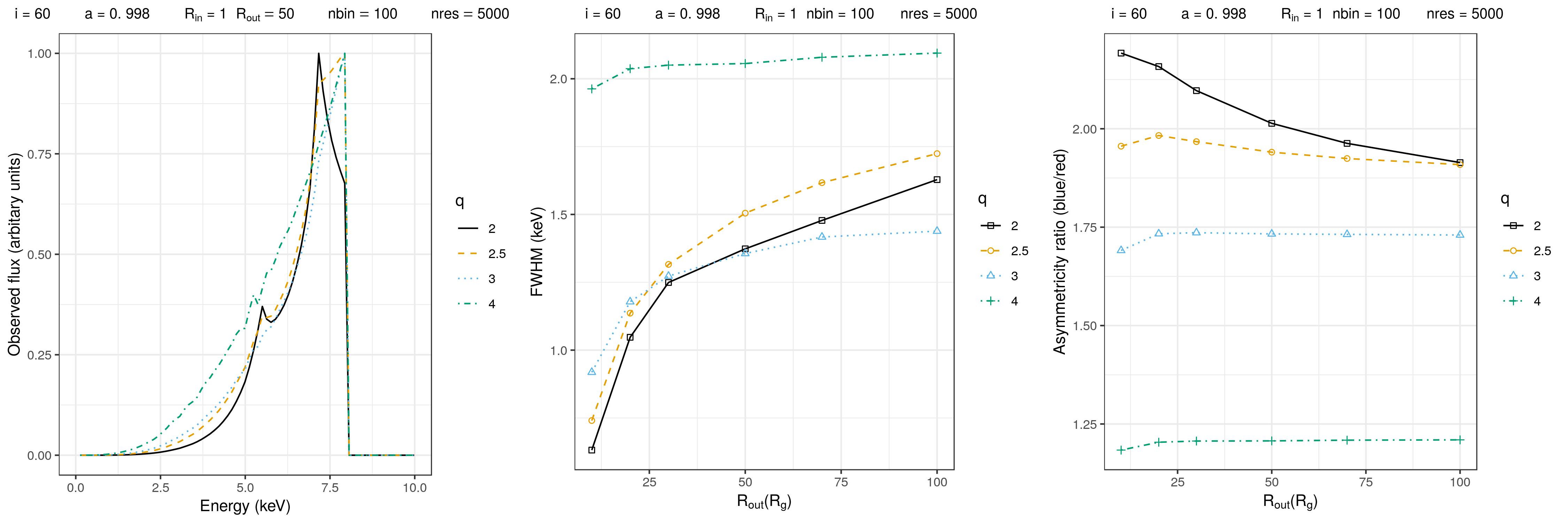} \\
\caption{Influence of power law emissivity index $q$ on the simulated line profiles (left panels),
its FWHM (middle panels) and asymmetricity ratio (right panels). The presented results correspond
to two different disc inclinations: $i=20^\circ$ and $i=60^\circ$, and SMBH spins: $a=0.005$
(practically non-rotating Schwarzschild SMBH) and $a=0.998$ (extremely rotating Kerr SMBH).}
\label{emis}
\end{figure*}

\section{Results}\label{sec:results}

\subsection{The effects of the spectral and spatial resolutions of the disc}

The obtained results show that during binning procedure one has to assume an 
appropriate number of line bins, since they could have significant effect on the 
resulting simulated line profiles. Namely, too small number of line bins will 
smooth the line profiles, and potentially hide some of the line's important 
features, such as its red peak (as demonstrated in the top left panel of Fig. 
\ref{nbin} for number of the bins less than $\approx 80$). Even in the case of 
highly inclined discs, when read peak is relatively strong (see second row of 
Fig. \ref{nbin}), its intensity and position could be affected by such 
smoothing. Besides, this smoothing can artificially increase asymmetricity ratio 
of the line profile (see the right panels of Fig. \ref{nbin}), and induce 
inaccuracies in its FWHM estimates (see the midle panels of the same figure), 
depending also on the spin of the central SMBH (as it can be seen by comparing 
the corresponding panels in the second and third row of Fig. \ref{nbin}). These 
effects are especially emphasized for higher values of emissivity index, since 
in this case it could also have significant influence on intensity and 
position of the blue peak (see the bottom row of Fig. \ref{nbin}).

The results obtained by simulation can be compared with the properties of the past, current and future X-ray detectors. It is known that the cameras of the XMM-Newton provide spectral resolving power $E/\Delta E \sim 20-50$ \cite{Turner2001,Struder}. The energy resolution of Suzaku satellite was 10 eV at 6 keV, and it provided a spectral resolving power $E/\Delta E \sim 600$ \cite{trumper,mitsuda}. For the future X-ray Integral Field Unit (X-IFU), that will be a part of the Athena X-ray Observatory planned energy resolution is $E/\Delta E \sim 2800$ in 0.2 - 12 keV range \cite{barret2016}.

In our simulation the energy resolution $E/\Delta E$ is taken to be in the range of the XMM-Newton. The energy resolution at 6.4 keV used in simulation is $E/\Delta E =$ 25, 35, 40 and 50, for $nbin = $ 50, 70, 80 and 100, respectively.

Regarding the number of photons received from the accretion disc ($nres\times 
nres$), in most cases it is sufficient to take $nres\approx 1000$, i.e. to 
collect $\propto 10^6$ of them in order to obtain the simulated Fe K$\alpha$ line 
profiles of with resonable quality, as it can be seen in the Fig. \ref{nres}. Only 
in the case of high emissivity index (see the bottom row of Fig. \ref{nres}) it 
is necessary to significantly increase the number of photons (i.e. the ''spatial 
resolution'' of the disc) in order to achieve this goal. 

The above results clearly demonstrate that both spectral and spatial 
resolutions of the X-ray detectors are of crucial significance for accurate 
measurements of FWHM and asymmetricity ratio in the observed Fe K$\alpha$ line 
profiles, and thus, for potential identification of these line profiles as 
relativistically broadened. In this paper we assumed spectral resolution which is similar to XMM-Newton resolution and investigated the influence of spectral resolution on the detection of relativistic Fe K$\alpha$ line in order to explore the ability of current detectors to observe (or not observe) this line. However, next generation of X-ray observatories (as e.g. ATHENA) will provide a higher spectral resolution (around 100 times better than current missions), and it is a task that we are going to explore (investigate) in a following paper.

\subsection{The effects of other disc parameters and SMBH spin}

Additionally, we show that the FWHM and asymmetry ratio of the observed Fe K$\alpha$ line profiles could be used for investigating the physics and geometry in the vicinity of SMBHs even with spectral resolution of current X-ray telescopes, and for this purpose  we simulated the
effects of the disc parameters and SMBH spin on these two quantities. The effects of SMBH spin $a$ on the simulated profiles of the Fe K$\alpha$ 
line, its FWHM and asymmetricity ratio are presented in Fig. \ref{spin}. As it 
can be seen  in Fig. \ref{spin}, the asymmetricity ratio increases, at first rapidly, 
with $R_{out}$ for the low disc inclination ($i=20^\circ$). For high disc 
inclination ($i=60^\circ$) asymmetricity ratio decrease with $R_{out}$ (see the 
right panel in the first and the second row, respectively). In the case of low 
emissivity index ($q=2$), asymmetricity ratio increase with $R_{out}<50$ and for 
the higher $R_{out}$ is almost constant, especially for high BH spin ($a>0.8$). For 
a high emissivity index ($q=4$), asymmetricity ratio decreses at first 
($R_{out}<25$) and after that it is constant as $R_{out}$ increases. In this case the 
asymmetricity ratio is almost constant for all values of $R_{out}$ for high 
BH spin ($a\ge 0.9$). For the low disc inclination ($i=20^\circ$), FWHM decreses 
with $R_{out}$ for $a\ge 0.4$ and increses for lower spins. In all cases FWHM 
$\approx 1.2$ for $R_{out}=20$ and starts to decrese for higher $R_{out}$ (see 
plot in the middle, first row). For higher inclinations FWHM increses with 
$R_{out}$. FWHM increses for $R_{out}<20$, independently of emissivity indexes, 
and becomes nearly constant for higher $R_{out}$. In the cases of hight BH spins 
($a\ge 0.9$), FWHM is nearly constant with $R_{out}$.

Fig. \ref{incl} shows influence of the disc inclination $i$ (i.e. viewing angle 
$\theta_{obs}$) on the line profile, FWHM and asymmetricity ratio. The 
presented results indicate that for lower disc inclinations ($i<40^\circ$) asymmetricity 
ratio increases with $R_{out}$ (see the right panels of Fig. \ref{incl}), for 
$i\approx 40^\circ$ it becomes nearly constant (especially for larger outer 
radii $R_{out}$), while for highly inclined discs ($i>40^\circ$) it decreases 
with $R_{out}$. This result implicates that asymmetricity ratio of the Fe 
K$\alpha$ line could be used for determining the outer radius of the line 
emitting region. 

Influence of power law emissivity index $q$ on the simulated line profiles, its 
FWHM and the asymmetricity ratio is presened in Fig. \ref{emis}, from which it can 
be seen that for all disc inclinations asymmetricity ratio increases, at first 
rapidly, with $R_{out}$. For high emissivity indexes ($q\ge 3$) asymmetricity 
ratio becomes nearly constant for $R_{out}>25$. The asymmetricity ratio is affected 
by SMBH spins in such a way that in the case of a non-rotating Schwarzschild SMBH 
($a=0.005$) the asymmetricity ratio is decresing with $R_{out}$. However, in the 
case of a rapidly rotating Kerr SMBH ($a=0.998$) the asymmetricity ratio is nearly 
constant with $R_{out}$ for emissivity indexes $q>2.5$. For the inclination 
$i=20^\circ$, FWHM increases rapidly for $R_{out}<20$. In tha cases of emissivity 
indexes $q\le3$, FWHM reaches the maximum at $R_{out}\approx 20$ and decreases 
as $R_{out}$ increases; however, for $q=4$ FWHM becomes almost constant for 
$R_{out}>20$. 

As it can be seen from Figs. \ref{nbin}-\ref{emis}, in most cases both, the FWHM 
and asymmetricity ratio of the Fe K$\alpha$ line strongly depend on disc outer 
radius $R_{out}$ and its inclination $i$.

\section{Conclusions}\label{sec:conclusions}

We developed a model of an accretion disc around SMBH hole using numerical 
simulations based on a ray-tracing method in the Kerr metric

This model allows us to study the radiation which originates in the vicinity of 
SMBHs. The shape of the emitted broad Fe K$\alpha$ line is strongly affected by 
three types of shifts: classical Doppler shift - causing double-peaked profile, 
special relativistic transverse Doppler shift and relativistic beaming - 
enhancing blue peak relative to red one and general relativistic gravitational 
redshift - smearing blue emission into red one.

Comparisons between the modelled and observed Fe K$\alpha$ line profiles allow 
us to determine the parameters of the line emitting region as well as to study 
plasma physics and spacetime metrics in
vicinity of SMBHs. Two of them are of an especial importance for the 
strong gravitational field investigation in AGN, i.e. the mass of central BH and its angular momentum. 
Other parameters can give us information about the plasma conditions in vicinity 
of the central BH of the AGN.

From our simulations, we find that  number of line bins and photons taken in calculations are of crucial significance for obtain the correct Fe K$\alpha$ line
profiles, especially in the case of higher the disc emissivity index. Also, the lack of observed Fe K$\alpha$ line can be caused by the low resolution (our bin simulation) and sensitivity (our number of photon simulation) of the X-ray detectors. In addition, we conclude that in most cases the FWHM and the asymmetricity ratio of the Fe K$\alpha$ line strongly depends on the parameters of the disc, especially the outer radius and inclination.

\section*{Acknowledgments}

This study is part of projects "Astrophysical Spectroscopy of Extragalactic 
Objects" (No. 176001), "Gravitation and the large scale structure of the 
Universe" (No. 176003) and "Visible and Invisible Matter in Nearby Galaxies: 
Theory and Observations” (No. 176021) supported by the Ministry of Education, 
Science and Technological development of Serbia. The work is partially supported 
by ICTP — SEENET-MTP project NT-03 "Cosmology - Classical and Quantum 
Challenges".

%\bibliographystyle{ws-ijmpa.bst}
%\bibliography{/home/mmilan/Documents/library}

\end{document}